\documentclass[aps,prd,twocolumn,showpacs,preprintnumbers,amsmath,amssymb,nofootinbib]{revtex4-1}
\usepackage{amsmath,amsfonts,euscript,amssymb}
\usepackage{colordvi}
\usepackage{color}
\usepackage{graphicx}
\usepackage{bm}
\DeclareSymbolFont{bfitletters}{OML}{cmm}{bx}{it}
\DeclareSymbolFont{bfitoperators}   {OT1}{cmr} {m}{n}
\DeclareMathSymbol{\bfitomega}{\mathord}{bfitletters}{"21}
\DeclareMathSymbol{\bfitgamma}{\mathord}{bfitletters}{"0D}
\newcommand{\be}{\begin{equation}}
\newcommand{\ee}{\end{equation}}
\newcommand{\bea}{\begin{eqnarray}}
\newcommand{\eea}{\end{eqnarray}}
\begin{document}

\title{Intrinsic time in Wheeler -- DeWitt conformal superspace}

\author{A. E. Pavlov$^{1,2}$}
\affiliation{${}^1$Bogoliubov~Laboratory~for~Theoretical~Physics,~Joint~Institute~of~Nuclear~Research,~
Joliot-Curie~str.~6,~Dubna,~141980,~Russia \\
$^{2}$Institute of Mechanics and Energetics, Russian State Agrarian University, Timiryazevskaya, 49,
Moscow 127550, Russia\\
alexpavlov60@mail.ru}

\date{\today}

\begin{abstract}
An intrinsic time in Geometrodynamics is obtained with using a scaled Dirac's mapping.
By addition of a background metric, one can construct a scalar field. It is suitable to play a role of
intrinsic time.
Cauchy problem was successfully solved in conformal variables because they are physical ones.
First, the intrinsic time as a logarithm of determinant of spatial metric, was applied to a cosmological
problem by Misner. A global time is exist under condition of constant mean curvature slicing of spacetime.
The volume of hypersurface and so-called mean York's time are canonical conjugated pair. So, the volume is the
intrinsic global time by its sense. The experimentally observed redshift in cosmology is the evidence of its
existence.
\end{abstract}

\maketitle

{\bf PACS} 04.20.Cv

\section{Introduction}

The Geometrodynamics is a theory of space and time in its inner essence.
The spatial metric $\gamma_{ij}$ carries information about the inner time.
The intrinsic time for cosmological models is constructed of inner metric characteristic of the space.
The time is to be a scalar relative to diffeomorphisms of changing coordinates of the space.
For this demand we use an idea of bimetric formalism, adding some auxiliary spatial metric.
Thus we naturally come to interpretations of observational data from the Conformal gravitation concepts.
The generalized Dirac's mapping \cite{Dirac} allows to extracting the intrinsic time.
The metric $\gamma_{ij}$ is factorized in the inner time factor $e^{-4D}$ and the conformal metric
$\tilde\gamma_{ij}$.

The Dirac's mapping reflects the transition to physical (conformal) variables.
In spirit of ideas of {\it Conformal cosmology} \cite{PP}, the conformal metric is a metric of the space,
where we live and make observations.
The choice of conformal measurement standards allows us to separate the cosmic evolution
of the devices of observation from the evolution of cosmic objects.
Thus we avoid an unpleasant artefact of expanding Universe and the inevitable problem of Big Bang in
Standard Cosmology. After the procedure of deparametrization is implemented the volume of the Universe occurs
its global intrinsic time. In modern papers \cite{SooYu, IIISooYu, Soo, Hoi, Eyo, Vasudev, Ramachandra, Shyam}
one can see applications of local intrinsic time interval in Geometrodynamics.

Werner Heisenberg in Chapter ``Quantum Mechanics and a Talk with Einstein (1925--1926)'' \cite{Heisenberg}
quoted Albert Einstein's statement:
``{\it But on principle, it is quite wrong to try founding a theory on observable magnitudes alone.
In reality the very opposite happens. It is the theory which decides what we can observe}''.
This conversation between two great scientists about the status of observable magnitudes in the theory
(quantum mechanics or General Relativity) remains actual nowadays.

\section{ADM variational functional of General Relativity. Notations}

The Einstein's General Relativity was presented in Hamiltonian form a half of century ago \cite{DiracProc}.
Paul Dirac manifested, that four-dimensional symmetry is not fundamental property of the physical world.
Instead of spacetime transformations, one should consider canonical transformations of the phase space
variables.

The ADM formalism, based on the Palatini approach,
was developed by Richard Arnowitt, Stanley Deser and Charles Misner in 1959 \cite{ADM}.
The formalism supposes the spacetime with interval
$$
{\bf g}=g_{\mu\nu}(t, {\bf x})dx^\mu\otimes dx^\nu
$$
is to foliated into a family of space-like surfaces $\Sigma_t,$ labeled by the {\it time coordinate} $t$,
and with spatial coordinates on each slice $x^i$. The metric tensor of spacetime in ADM form looks like
\be\label{gmunu} (g_{\mu\nu})=
\left(
\begin{array}{cc}
-N^2+N_iN^i& N_i\\
N_j& \gamma_{ij}
\end{array}
\right).
\ee
The physical meaning of the metric components are the following:
the lapse function $N(t; x,y,z)$ defines increment of coordinate time $t$, and the shift vector
$N_i (t; x,y,z)$ defines replacement of coordinates of hypersurface under transition to an infinitesimally
close spacetime hypersurface.

The first quadratic form
\be\label{spacemetric}
{\bfitgamma}=\gamma_{ik}(t, {\bf x})dx^i\otimes dx^k
\ee
defines the induced metric on every slice $\Sigma_t$.
The components of spatial matrix $\gamma_{ij}(t; x,y,z)$ (\ref{spacemetric})
contain three gauge functions
describing the spatial coordinates. The three remaining components describe two polarizations of gravitational
waves and many-fingered time. So, we defined the foliation $(\Sigma_t, \gamma_{ij})$.

The group of general coordinate transformations conserving such a foliation was found by Abraham Zel'manov
\cite{Zel}, this group involves the reparametrization subgroup
of coordinate time.
This means that the coordinate time, which is not invariant with respect to gauges, in general case,
is not observable.
A large number of papers were devoted to the choice of reference frames (see, for example, a monograph
\cite{Vlad} and references therein).

The components of the extrinsic curvature tensor $K_{ij}$ of every slice are constructed out of the second
quadratic form of the hypersurface, and can be defined as
\be\label{Kij}
K_{ij}:=-\frac{1}{2}{\pounds}_{\bf n}\gamma_{ij},
\ee
where ${\pounds}_{\bf n}$ denotes the Lie derivative along the $n^\mu$, a time-like unit normal to the slice,
direction.
The components of the extrinsic curvature tensor can be found by the formula
\be\label{Kijdef}
K_{ij}=\frac{1}{2N}\left(\nabla_i N_j+\nabla_j N_i-\dot\gamma_{ij}\right),
\ee
where $\nabla_k$ is a Levi--Civita connection associated with the metric $\gamma_{ij}$
$$\nabla_k \gamma_{ij}=0.$$

The Hamiltonian dynamics of General Relativity is built in an infinite-dimensional degenerated phase space
of 3-metrics $\gamma_{ij} ({\bf x},t)$ and densities of their momenta $\pi^{ij}({\bf x},t)$.
The latter are expressed through the tensor of extrinsic curvature
\be\label{piij}
\pi^{ij} := -\sqrt{\gamma}(K^{ij}-K\gamma^{ij}),
\ee
where we introduced notations:
\bea
K^{ij}&:=&\gamma^{ik}\gamma^{jl}K_{kl},\qquad K:=\gamma^{ij}K_{ij},\nonumber\\
\gamma&:=&\det||\gamma_{ij}||,\qquad \gamma_{ij}\gamma^{jk}=\delta_i^k.\label{gammadet}
\eea
The Poisson bracket is a bilinear operation on
two arbitrary functionals $F[\gamma_{ij}, \pi^{ij}],$ $G[\gamma_{ij}, \pi^{ij}]$  \cite{T}
\bea\label{PB}
&&\{F, G\}=\\
&&\int\limits_{\Sigma_t}\, d^3x\left(\frac{\delta F}{\delta \gamma_{ij}(t,{\bf x})}
\frac{\delta G}{\delta \pi^{ij}(t,{\bf x})}- \frac{\delta G}{\delta \gamma_{ij}(t,{\bf x})}
\frac{\delta F}{\delta \pi^{ij}(t,{\bf x})}\right).\nonumber
\eea

The canonical variables satisfy to the relation
\be\label{gammapi}
\{\gamma_{ij}(t,{\bf x}),\pi^{kl}(t,{\bf x}')\}=\delta_{ij}^{kl}\delta ({\bf x}-{\bf x}'),
\ee
where
$$
\delta_{ij}^{kl}:=\frac{1}{2}\left(\delta_i^k\delta_j^l+\delta_i^l\delta_j^k\right),
$$
and $\delta ({\bf x}-{\bf x}')$ is the Dirac's $\delta$-function for the volume of $\Sigma_t$.

The super-Hamiltonian of the gravitational field is a functional
\be\label{superHam}
\int\limits_{\Sigma_t}
\left( N{\cal H}_\bot+N^i{\cal H}_i\right)\, d^3x,
\ee
where $N$ and $N^i$ are Lagrange multipliers, ${\cal H}_\bot$, and ${\cal H}_i$ have sense of constraints.
Among them,
\be\label{constraintHam}
{\cal H}_\bot :=
G_{ijkl}\pi^{ij}\pi^{kl}-\sqrt{\gamma}R (\gamma_{ij})
\ee
is obtained from the scalar Gauss relation of the embedding hypersurfaces theory and
called the Hamiltonian constraint. Here $R$ is the Ricci scalar of the space,
$$
 G_{ijkl}:=\frac{1}{2\sqrt{\gamma}}(\gamma_{ik}\gamma_{jl}+\gamma_{il}\gamma_{jk}-\gamma_{ij}\gamma_{kl})
$$
is the supermetric of the 6-dimensional hyperbolic Wheeler -- DeWitt (WDW) superspace \cite{DeWitt}.
Momentum constraints
\be\label{constraintMom}
{\cal H}^i := -2\nabla_j\pi^{ij}
\ee
are obtained from the contracted Codazzi equations of the embedding hypersurfaces theory. They
impose restrictions on possible data $\gamma_{ij}({\bf x},t), \pi^{ij}({\bf x},t)$
on a space-like hypersurface
$\Sigma_t$.
The divergence law, following from (\ref{constraintMom}), is analogous to the Gauss law in Maxwell's
electrodynamics. The Hamiltonian constraint (\ref{constraintHam}) has no analogue in  electrodynamics.
It yields the dynamics of the space geometry itself.
The Hamiltonian dynamics is built of the ADM - variational functional
\be\label{ADMvar}
W=\int\limits_{t_I}^{t_0}
\, dt\int\limits_{\Sigma_t}
d^3x\left(\pi^{ij}\frac{d\gamma_{ij}}{dt}-N{\cal H}_\bot-N^i{\cal H}_i\right),
\ee
where ADM units: $c=1, 16\pi G=1$ were used \cite{ADM}.
The action (\ref{ADMvar}) is obtained of the Hilbert functional after the procedure of $(3 + 1)$ foliation
and the Legendre transformation executed.

These constraints are of the first class, because they identify to the closed algebra
\begin{eqnarray}
&&\{{\cal H}_\bot (t,{\bf x}), {\cal H}_\bot (t,{\bf x}')\}=\nonumber\\
&&({\cal H}^i (t,{\bf x})+{\cal H}^i (t,{\bf x}'))
\delta_{,i}({\bf x}-{\bf x}'),\nonumber\\
&&\{{\cal H}_i (t,{\bf x}), {\cal H}_\bot (t,{\bf x}')\}={\cal H}_\bot (t,{\bf x})
\delta_{,i}({\bf x}-{\bf x}'),
\nonumber\\
&&\{{\cal H}_i (t,{\bf x}), {\cal H}_j (t,{\bf x}')\}=\nonumber\\
&&{\cal H}_i (t,{\bf x}')
\delta_{,j}({\bf x}-{\bf x}')+ {\cal H}_j (t,{\bf x})\delta_{,i}({\bf x}-{\bf x}').\nonumber
\end{eqnarray}
The Poisson brackets between constraints vanish on the constraints hypersurface.
In the presence of matter, described by the energy-momentum tensor $T_{\mu\nu}$, the considered constraints
(\ref{constraintHam}), (\ref{constraintMom}) take the form of Einstein's equations
\be\label{Ham0}
{\cal H}_\bot=\sqrt{\gamma}\left(K_{ij}K^{ij}-K^2\right)-\sqrt{\gamma}R+\sqrt{\gamma} T_{\bot\bot},
\ee
\be\label{mom0}
{\cal H}^i=-2\sqrt{\gamma}\nabla_j\left(K^{ij}-\gamma^{ij}K\right)+\sqrt{\gamma} (T_\bot)^i,
\ee
where
\be\label{Tbotbot}
T_{\bot\bot}:=n^\mu n^\nu T_{\mu\nu}
\ee
is the matter density, and
\be\label{Tboti}
(T_\bot)_i:= n^\mu T_{i\mu}
\ee
is the matter momentum density in a normal observer (Euler observer) reference.
The Hamiltonian constraint (\ref{Ham0}) can be expressed in the momentum variables (\ref{piij}):
\be\label{Hc}
{\cal H}_\bot=\frac{1}{\sqrt\gamma}\left(\pi_{ij}\pi^{ij}-\frac{1}{2}\pi^2\right)-\sqrt\gamma R+
\sqrt\gamma T_{\bot\bot},
\ee
as far as
$$
K^{ij}=-\frac{1}{\sqrt{\gamma}}\left(\pi^{ij}-\frac{1}{2}\pi\gamma^{ij}\right),\qquad
\pi_{ij}:=\gamma_{ik}\gamma_{jl}\pi^{kl},
$$
$$\pi:=\gamma_{ij}\pi^{ij},\qquad K=\frac{\pi}{2\sqrt{\gamma}}.
$$
The momentum constraints (\ref{mom0}) in the momentum variables are the following:
\be\label{mc}
{\cal H}^i=-2\nabla_j\pi^{ij}+\sqrt\gamma (T_\bot)^i.
\ee

The Poisson structure (\ref{gammapi}) is degenerated by force of existence of the constraints
(\ref{Hc}), (\ref{mc}).
The reduction of the dynamical system on the level of the constraints, in general case, is an open problem.

\section{Shape dynamics}

A.A. Friedmann in his book \cite{Friedmann}, dedicated to cosmology of the Universe, found the following
remarkable words about the principle of scale invariance:
``{\it ... moving from country to country, we have to change the scale, id est, measured in Russia --
by arshins, Germany -- meters, England -- feet. Imagine that such a change of scale we had to do from point
to point, and then we got the above operation of changing of scale. Scale changing in the geometric world
corresponds, in the physical world, to different ways of measuring of the length... Properties of the world,
are divided into two classes: some are independent of the above said change of scale, better to say, do not
change their shape under any changes of scale, while others under changing of the scale, will change their
shape. Let us agree on their own properties of the world, belonging to the first class, and call scale
invariant. Weyl expands the invariance postulate, adding to it the requirements that all physical laws were
scale-invariant properties of the physical world. Consistent with such an extension of the postulate of
invariance, we have to demand that the world equations would be expressed in a form satisfactory to not only
coordinate, but the scale invariance}''.
Radiative breaking of conformal symmetry in a conformal-invariant version of the Standard Model of elementary particles is considered in \cite{Euro}. The fruitful idea of initial conformal symmetry of the theory lead to 
right value of Higgs boson mass.

The Einstein's theory of general relativity is covariant under general coordinate transformations.
The group of transformations is an infinite-parameter one. The action of the group can be reduced to
alternating actions of its two finite-parameter subgroups: the spatial linear group $SL (3,1)$ and the
conformal group $SO(4,2)$.
According to Ogievetsky's theorem \cite{Og}, the invariance under the infinite-parameter generally covariant
group is equivalent to simultaneous invariance under the affine and the conformal group.
Using an analogy with phenomenological chiral Lagrangians \cite{VP}, it is possible to obtain phenomenological
affine Lagrangian as nonlinear joint realization of affine and conformal symmetry groups. A nonlinear
realization  of the affine group leads to a symmetry tensor field as a Goldstone field. The requirement that
the theory correspond simultaneously to a realization of the conformal group as well leads uniquely to the
theory of a tensor field whose equations are Einstein's  \cite{BO}. York's method of decoupling of the
momentum and Hamiltonian constraints \cite{York} is derived on a basis of a mathematical discovery.
The physical principle of York's method is based on initially being conformal and affine symmetry of the theory. that is proved in paper \cite{BO}.

For recovering an initial conformal symmetry to the space one uses an artificial method.
One can possible to change the gauge symmetry of General Relativity (spatial diffeomorphisms and local
changing of slicing) by gauge symmetry of dynamics of form (spatial diffeomorphisms and local scaling conserved
the global slicing) of spacetime \cite{Foster}.
Following to \cite{Foster}, let us define a class of metrics $\bfitgamma$ of some hypersurface $\Sigma$,
that conserve its volume
$$
{V_\gamma}=\int_\Sigma\, d^3x\sqrt{\gamma (x)},
$$
with help of conformal mapping of metric coefficients
\be
\label{conformalmetric}
\gamma_{ij}(x)\to\exp (4\hat{\phi}(x))\gamma_{ij}(x).
\ee
Here the function is defined
\begin{equation}\label{hatphi}
\hat{\phi}({x}):=\phi(x)-\frac{1}{6}\ln <e^{6\phi}>_{\gamma},
\end{equation}
and operation of meaning by a hypersurface $\Sigma$ for some scalar field $f$
\begin{equation}
<f>_\gamma:=\frac{1}{V_\gamma}\int_\Sigma\,d^3x\sqrt{\gamma (x)}f(x).
\end{equation}
There was introduced St$\ddot{\rm u}$ckelberg scalar field \cite{Stuk} in space by analogy with Deser's
\cite{Deser} introducing of dilaton Dirac field \cite{Diracdilaton} in spacetime and an averaging of
functions \cite{ABNPBPZ,BPZZ} for an arbitrary manifold.

The theorem: {\it conformal mapping (\ref{conformalmetric}) conserves a volume of every hypersurface} was
proved in \cite{Gomes}. From the definition (\ref{hatphi}) the conformal factor is expressed
\be
e^{4\hat\phi}=\frac{e^{4\phi}}{\left[<e^{6\phi}>_\gamma\right]^{2/3}}.
\ee
Then Jacobian of transformation is transformed by the formula
$$
\sqrt\gamma\to\frac{e^{6\phi}}{<e^{6\phi}>_\gamma}\sqrt\gamma.
$$
The variation of Jacobian and, correspondingly, of the volume of hypersurface are
$$
\delta\sqrt\gamma=\frac{1}{2}\sqrt\gamma\gamma^{ab}\delta\gamma_{ab},\quad
{\delta V_\gamma}=\frac{1}{2}\sqrt{\gamma (y)}\gamma^{ij}(y){\delta\gamma_{ij}(y)}.
$$
The volume of the hypersurface $V_\gamma$ is conserved
\bea
V_\gamma&=&\int_{\Sigma_t}d^3x\,\sqrt{\gamma}(x)\to\frac{1}{<e^{6\phi}>_\gamma}
\int_{\Sigma_t}d^3x\,e^{6\phi}\sqrt{\gamma}(x)\nonumber\\
&=&V_\gamma\frac{<e^{6\phi}>_\gamma}{<e^{6\phi}>_\gamma}=V_\gamma.\nonumber
\eea
Q.E.D.

The phase space (cotangent bundle over $Riem (\Sigma)$) can be extended with the scalar field $\phi$ and
canonically conjugated momentum density $\pi_\phi$.
Let ${\cal C}$ is a group of conformal transformations of the hypersurface $\Sigma$:
$$(\gamma_{ij},\pi^{ij}; \phi,\pi_\phi)\mapsto (\Gamma_{ij},\Pi^{ij}; \Phi,\Pi_\phi),$$
parameterized by scalar field $\phi$ with a generation functional \cite{Gomes}
\bea\label{genF}
&&F_\phi [\gamma_{ij}, \Pi^{ij}, \phi, \Pi_\phi]:=\\
&&\int_\Sigma\, d^3x\left[\gamma_{ij}({x})e^{4\hat{\phi}({x})}\Pi^{ij}({x})+\phi({x})\Pi_\phi({x})\right].\nonumber
\eea

The canonical transformations in the extended phase space
$$(\gamma_{ij},\pi^{ij};\phi,\pi_\phi)\in\Gamma_{Ext}:=\Gamma\times T^*({\cal C})$$
with the canonical Poisson bracket are the following:
\bea
&&\gamma_{ij}\to{\cal T}_\phi\gamma_{ij}(x):=\Gamma_{ij}=e^{4\hat\phi (x)}\gamma_{ij},\label{can1}\\
&&\pi^{ij}\to{\cal T}_\phi\pi^{ij}(x):=\Pi^{ij}=\label{can2}\\
&&e^{-4\hat\phi (x)}\left(
\pi^{ij}(x)-\frac{1}{3}\gamma^{ij}\sqrt\gamma <\pi>(1-e^{6\hat\phi})\right),\nonumber\\
&&\phi (x)\to{\cal T}_\phi \phi (x):=\Phi=\phi (x),\label{can3}\\
&&\pi_\phi(x)\to{\cal T}_\phi\pi_\phi (x):=\Pi_\phi=\label{can4}\\
&&\pi-4\left(\pi_\phi (x)-<\pi>\sqrt\gamma\right).\nonumber
\eea

After the canonical transformations were implemented, the constraints (\ref{Ham0}), (\ref{mom0})
got the following form:
\bea
\left({\cal H}_\bot\right)_\phi&=&
\frac{e^{-6\hat\phi}}{\sqrt\gamma}
\left(\pi^{ij}\pi_{ij}+\frac{1}{3}\sqrt\gamma\left(1-e^{6\hat\phi}\right)
<\pi>\pi-\right.\nonumber\\
&-&\left.
\frac{1}{6}\gamma\left(1-e^{6\hat\phi}\right)^2<\pi>^2-\frac{1}{2}\pi^2\right)-\nonumber\\
&-&\sqrt\gamma\left(Re^{2\hat\phi}-8e^{\hat\phi}\Delta e^{\hat\phi}\right),\nonumber\\
\left({\cal H}^i\right)_\phi&=&
-2e^{-4\hat\phi}\left(\nabla_j\pi^{ij}-
2(\pi-\sqrt\gamma <\pi>)\nabla^i\phi\right),\nonumber\\
{\cal Q}_\phi&=&\pi_\phi-4(\pi-<\pi>\sqrt\gamma).\nonumber
\eea

\section{Cauchy problem in Conformal gravitation}

Let us proceed the solution of the Cauchy problem following to York in conformal variables (denoted by bar)
in detail. ``{\it Note that the configuration space that one is led to by the initial-value equations is not
superspace (the space of Riemannian three-geometries), but ``conformal superspace'' [the space of which each
point is a conformal equivalence class of Riemannian three-geometries]$\times$[the real line]
(i.e., the time $T$)}'' \cite{York}.

The matter characteristics under conformal transformation
\begin{equation}\label{Psifactor}
\gamma_{ij}:=e^{-4\hat\phi}\bar\gamma_{ij}\equiv\Psi^4\bar\gamma_{ij}
\end{equation}
are transformed according to their conformal weights. We denote the transformed matter characteristics
(\ref{Tbotbot}) and (\ref{Tboti}) as
\begin{equation}\label{tildeT}
\bar{T}_{\bot\bot}:=\Psi^8 T_{\bot\bot};\qquad
(\bar{T}_\bot)^i:=\Psi^{10}({T}_\bot)^i.
\end{equation}

After the traceless decomposition of $K^{ij}$
\be\label{KijAijK}
K^{ij}=A^{ij}+\frac{1}{3}K\gamma^{ij},\qquad \gamma_{ij}A^{ij}=0,
\ee
we decompose the traceless part of $A^{ij}$ according to
\be\label{Aijhat}
A^{ij}=\Psi^{-10}\bar{A}^{ij}.
\ee
Then, we obtain conformal variables
$$
\bar{A}^{ij}:=\Psi^{10}A^{ij},\qquad \bar{A}_{ij}:=\bar\gamma_{ik}\bar\gamma_{jl}\bar{A}^{kl}=
$$
$$=
\Psi^{-8}\gamma_{ik}\gamma_{jl}\bar{A}^{kl}=\Psi^2 A_{ij}.
$$
The Hamiltonian constraint (\ref{Ham0}) in the new variables
\bea\nonumber
&&\bar\Delta\Psi-\frac{1}{8}\bar{R}\Psi+\frac{1}{8}\bar{A}_{ij}\bar{A}^{ij}\Psi^{-7}-\nonumber\\
&&-\frac{1}{12}K^2\Psi^5+
\frac{1}{8}\bar{T}_{\bot\bot}\Psi^5=0
\label{H}
\eea
is named the Lichnerowicz -- York equation \cite{LYork}.
Here $\bar\Delta:=\bar\nabla_i\bar\nabla^i$ is the conformal Laplacian,
$\bar\nabla_k$ is the conformal connection associated with the conformal metric $\bar\gamma_{ij}$
$$\bar\nabla_k\bar\gamma_{ij}=0,$$
$\bar{R}$ is the conformal Ricci scalar expressed of the Ricci scalar $R$:
\be\label{Rscalar}
R= \Psi^{-4}\bar{R}-8 \Psi^{-5}\bar\Delta\Psi.
\ee
Lichnerowicz originally considered the differential equation (\ref{H})
without matter and in case of a maximal slicing gauge $K=0$ \cite{Lich}.

The momentum constraints (\ref{mom0}) after the decomposition (\ref{Aijhat}) take the form:
\be\label{M}
\bar\nabla_j\bar{A}^{ij}-\frac{2}{3}\Psi^6\bar\nabla^i K+\frac{1}{2}(\bar{T}_\bot)^i=0.
\ee

To solve the Cauchy problem in the General Relativity, York elaborated the conformal
transverse -- traceless method
\cite{YorkSources}.
He made the following decomposition of the traceless part $\bar{A}_{ij}$:
\be\label{longtrans}
\bar{A}^{ij}=\left(\bar{\mathbb L}{X}\right)^{ij}+\bar{A}_{TT}^{ij},
\ee
where $\bar{A}_{TT}^{ij}$ is both traceless and transverse with respect to the metric $\bar\gamma_{ij}$:
$$\bar\gamma_{ij}\bar{A}_{TT}^{ij}=0,\qquad \bar\nabla_j\bar{A}_{TT}^{ij}=0,$$
$\bar{\mathbb L}$ is the {\it conformal Killing operator}, acting on the vector field ${\bf X}$:
\be\label{Killing}
\left(\bar{\mathbb L} {X}\right)^{ij}:=\bar\nabla^i X^j+\bar\nabla^j X^i-
\frac{2}{3}\bar\gamma^{ij}\bar\nabla_k X^k.
\ee
The symmetric tensor $\left(\bar{\mathbb L} {X}\right)^{ij}$ is called the {\it longitudinal part} of
$\bar{A}^{ij}$, whereas
$\bar{A}_{TT}^{ij}$ is called the {\it transverse part} of $\bar{A}^{ij}$.

Using the York's longitudinal-transverse decomposition (\ref{longtrans}), the constraint equation (\ref{H})
can be rewritten in the following form
\bea
&&\bar\Delta\Psi-\frac{1}{8}\bar{R}\Psi+\nonumber\\
&+&\frac{1}{8}\left[\left(\bar{\mathbb L} {X}\right)_{ij}+
\bar{A}_{ij}^{TT}\right]\left[\left(\bar{\mathbb L} {X}\right)^{ij}+
\bar{A}_{TT}^{ij}\right]\Psi^{-7}-\nonumber\\
&-&\frac{1}{12}K^2\Psi^5+
\frac{1}{8}\bar{T}_{\bot\bot}\Psi^5=0,\label{HYork}
\eea
where the following notations are utilized
$$\left(\bar{\mathbb L} {X}\right)_{ij}:=
\bar\gamma_{ik}\bar\gamma_{jl}\left(\bar{\mathbb L} {X}\right)^{kl},\qquad
\bar{A}_{ij}^{TT}:=\bar\gamma_{ik}\bar\gamma_{jl}\bar{A}_{TT}^{kl};$$
and the momentum equations (\ref{M}) are:
\be\label{MYork}
\bar\Delta_{\mathbb L}X^i-\frac{2}{3}\Psi^6\bar\nabla^i K + \frac{1}{2}(\bar{T}_\bot)^i=0.
\ee
The second order operator $\bar\nabla_j\left(\bar{\mathbb L} {X}\right)^{ij}$, acting on the vector
${\bf X}$, is the
{\it conformal vector Laplacian} $\bar\Delta_{\mathbb L}$:
\bea
\bar\Delta_{\mathbb L} X^i&:=&\bar\nabla_j\left(\bar{\mathbb L} X\right)^{ij}=\nonumber\\
&=&\bar\nabla_j\bar\nabla^j X^i+\frac{1}{3}\bar\nabla^i\bar\nabla_j X^j+\bar{R}^i_j X^j.\label{Laplacian}
\eea
To obtain the formula (\ref{Laplacian}), we have used the contracted Ricci identity.

The part of the initial data on $\Sigma_0$ can be freely chosen and other part is constrained, {\it id est}
determined from the
constrained equations (\ref{HYork}), (\ref{MYork}).
One can offer a constant mean curvature condition on Cauchy hypersurface $\Sigma_0$:
\be\label{piT}
K\equiv \frac{\pi}{2\sqrt{\gamma}}=\rm{const}.
\ee
Then the momentum constraints (\ref{mom0}) are separated of the Hamiltonian constraint (\ref{Ham0}) and reduce
to
\be\label{Poisson}
\bar\Delta_{\mathbb L}X^i+\frac{1}{2} (\bar{T}_\bot)^i=0.
\ee
Therefore, we obtain the {\it conformal vector Poisson equation}.
It is solvable for closed manifolds, as it was proved in \cite{ChBIYork}. So, we have

$\bullet$
Free initial data:

conformal factor $\Psi^4$;
conformal metric $\bar\gamma_{ij}$;
transverse tensor $\bar{A}_{TT}^{ij}$;
conformal matter variables $(\bar{T}_{\bot\bot}, (\bar{T}_\bot)^i)$.

$\bullet$
Constrained data:

scalar field $K$;
vector ${\bf X}$, obeying the linear elliptic equations (\ref{MYork}).

Note, after solving the Cauchy problem, we are not going to return to the initial variables,
in opposite to the York's approach. Cauchy problem was successfully solved not by chance after mathematically
formal transition to conformal variables. The point is that we have found just the physical variables.

According to Yamabe's theorem \cite{Yamabe}, {\it any metric of compact Riemannian manifold of dimension more
or equal three, can be transformed to a metric of space with constant scalar curvature.}
If we multiply the equation (\ref{Rscalar})
to $\Psi^{-1}$ and integrate it over all manifold $\Sigma$:
\bea
&&\int_\Sigma\,d^3x\sqrt\gamma R\Psi^{-1}=\int_\Sigma\, d^3x\sqrt{\bar\gamma}(\bar{R}\Psi-8\bar\Delta\Psi )
\equiv\nonumber\\
&&\int_\Sigma\,d^3x\sqrt{\bar\gamma}\bar{R}\Psi.\nonumber
\eea
For compact manifolds, the integral of Laplacian of the scalar function is equal to zero. Consequently,
a sign of scalar of curvature is conserved under conformal transformations. There is a Yamabe's constant
\be
{\textsf{y}}[\Sigma,\gamma ]=\inf_\Psi\left\{\frac{\int d^3x\sqrt\gamma (\Psi^2 R-8\Psi\Delta\Psi )}
{\int d^3x\sqrt\gamma \Psi^6}\right\}.
\ee
According to Yamabe's theorem, in a class of conformal equivalent class of metrics there is metrics,
where a minimum is realized, and hence, they correspond to spaces of constant scalar curvature.
Riemannian metrics are classified according to a sign of Yamabe constant. They are classified to positive,
negative and zeroth ones.

Multiplying the Lichnerowicz -- York equation (\ref{H}) to $8\Psi^7$, rewrite it in the following form
\be\label{LYm}
8\Psi^7\bar\Delta\Psi=\left(\frac{2}{3}K^2-\bar{T}_{\bot\bot}\right)\Psi^{12}+\bar{R}\Psi^8-
\bar{A}_{ij}\bar{A}^{ij}.
\ee
Then we denote
$z\equiv\Psi^4$
and present the right side of the equation (\ref{LYm}) as a polynomial
\be\label{polynomial}
f(z)=\left(\frac{2}{3}K^2-\bar{T}_{\bot\bot}\right) z^3+\bar{R} z^2-\bar{A}_{ij}\bar{A}^{ij}.
\ee
According to the theory of quasilinear elliptic equations, the equation
has a unique solution, if the polynomial of the third order by $z$ (\ref{polynomial})
$$f(z)=\left(\frac{2}{3}K^2-\bar{T}_{\bot\bot}\right)(z-z_1)(z-z_2)(z-z_3)$$
has a unique real root $z_i$.
A generic solution of the York's problem of initial problem is valid in spacetime near the
Cauchy hypersurface of constant mean curvature $(CMC).$
We restrict our consideration here by matter sources adopted to the theorem of existence.

\section{Many-fingered intrinsic time in Geometrodynamics}

Dirac, in searching of dynamical degrees of freedom of the gravitational field, introduced {\it conformal
field variables} $\tilde\gamma_{ij}, \tilde\pi^{ij}$ \cite{Dirac}
\begin{equation}\label{hatgamma}
\tilde\gamma_{ij}:=\frac{\gamma_{ij}}{\sqrt[3]{\gamma}}, \qquad
\tilde\pi^{ij}:=\sqrt[3]{\gamma}\left(\pi^{ij}-\frac{1}{3}\pi\gamma^{ij}\right),
\end{equation}
id est, our choice of the conformal factor (\ref{Psifactor}):
\be\label{fixshape}
\Psi=\gamma^{1/12}.
\ee
Among them, there are only five independent pairs $(\tilde\gamma_{ij}, \tilde\pi^{ij})$ per space point,
because
of unity of the determinant of conformal metric and traceless of the matrix of conformal momentum densities
$$\tilde\gamma:=\det ||\tilde\gamma_{ij}||=1,\qquad \tilde\pi:=\tilde\gamma_{ij}\tilde\pi^{ij}=0.$$
The remaining sixth pair $(D, \pi_D)$
\begin{equation}\label{hatTpT}
D:=-\frac{1}{3}\ln\gamma,\qquad \pi_D:=\pi
\end{equation}
is canonically conjugated. The essence of the transformation (\ref{hatgamma}) lies in the fact, that
the metric $\tilde\gamma_{ij}$ is equal to the whole class of conformally equivalent Riemannian
three-metrics $\gamma_{ij}$. So, the conformal variables describe dynamics of shape of the hypersurface of
constant volume.
The conformal mapping is not a coordinate diffeomorphism. Under the conformal transformation any
angles between vectors are equal and ratios of their
magnitudes are preserved in points with equal coordinates of the spaces \cite{Eisenhart}.
The extracted canonical pair $(D, \pi_D)$ (\ref{hatTpT}) has a transparent physical sense, {viz}:
an intrinsic time $D$ and a Hamiltonian density of gravitation field $\pi_D$.

The concepts of
{\it Quantum Geometrodynamics} of Bryce DeWitt \cite{DeWitt} and John Wheeler \cite{Wheeler}
follow from the Dirac's transformations (\ref{hatgamma})--(\ref{hatTpT}).
They thought the cosmological time to be identical the cosmological scale factor.
Information of time must be contained
in the internal geometry, and Hamiltonian must be given by the characteristics of external geometry (\ref{Kij}).
The canonical variable $D$ (\ref{hatTpT}) plays a role of intrinsic time in Geometrodynamics.

However, generally, as we see from (\ref{hatgamma})-(\ref{hatTpT}), $D$ is not a scalar,
and $\tilde\gamma_{ij}$ is not a tensor under group of diffeomorphisms.
According to terminology in \cite{Zel}, $D$ is not a kinemetric invariant,
$\tilde\gamma_{ij}$ is not a kinemetric invariant 3-tensor under kinemetric group of transformation.
The Dirac's transformations (\ref{hatgamma})--(\ref{hatTpT}) have a limited range of applicability:
they can be used in the coordinates with dimensionless metric determinant.

To construct physical quantities, an approach based on Cartan differential forms, invariant under
diffeomorphisms has been elaborated \cite{ABNPBPZ}.
In the present paper, to overcome these difficulties we use a fruitful idea of {\it bimetric formalism}
\cite{Rosen}.
Spacetime bi-metric theories were founded on some auxiliary background non - dynamical metric
with coordinate components
$f_{ij}({\bf x})$ of some 3-space, Lie-dragged along the coordinate time evolution vector
\begin{equation}\label{dfij}
\frac{\partial f_{ij}}{\partial t}=0.
\end{equation}
A background flat space metric was used for description of asymptotically flat spaces.
It was connected with Cartesian coordinates which are natural for such problems.
The development of these theories has been initiated by the problem of energy in Einstein's
theory of gravitation. The restriction (\ref{dfij}) to the background metric is not strong from mathematical
possibilities, so there is a possibility of its choosing from physical point of view.
If a topology of the space is $S^3$ so the background metric is a metric of sphere; for $S^2\times S^1$ --
the metric corresponding to Hopf one; and for $S^1\times S^1\times S^1$ the metric of flat space should be taken.
The space bi-metric approach is natural for solution of cosmological
problems also, when a static tangent space will be taken in capacity of a background space.

To use an auxiliary metric for a generic case of spatial manifold with arbitrary topology,
let us take a local tangent space  ${\cal T}(\Sigma_t)_{\bf x}$ as a background space
for every local region of our manifold $(\Sigma_t)$.
In every local tangent space we define a set of three linear independent vectors $e^i_{a}$ (dreibein),
numerated by the first Latin indices $a, b$.
The components of the background metric tensor in the tangent space:
$$e^i_{a}e_{b i}=f_{a b}.$$
Along with the dreibein, we introduce three mutual vectors $e^{a}_i$ defined by the orthogonal conditions
$$e_i^{a}e^i_{b}=\delta_b^a,\qquad e_i^{a}e^j_{a}=\delta_i^j.$$
Then, we can construct three linear independent under diffeomorphisms Cartan forms \cite{VP}
\begin{equation}\label{dif}
\omega^{a}(d)=e^{a}_i dx^i.
\end{equation}
The background metric is defined by the differential forms (\ref{dif}):
\begin{equation}
{\bf f}=f_{a b}\omega^{a}(d)\otimes\omega^{b}(d)=f_{ij}dx^i\otimes dx^j,
\end{equation}
where the components of the background metric tensor in the coordinate basis are
$$f_{ij}=f_{a b}e^{a}_i e^{b}_j.$$
The components of the inverse background metric denoted by $f^{ij}$ satisfy to the condition
$$f^{ik}f_{kj}=\delta^i_j.$$
Now, we can compare the background metric with the metric of gravitational field in every
point of the manifold $\Sigma_t$ in force of biectivity of mapping
$$\Sigma_t \longleftrightarrow {\cal T}(\Sigma_t)_{\bf x}.$$
The Levi--Civita connection $\bar\nabla_k$ is associated with the background metric $f_{ij}$:
$$\bar\nabla_k f_{ij}=0.$$

Let us define {\it scaled Dirac's conformal variables} $(\tilde\gamma_{ij}, \tilde\pi^{ij})$
by the following formulae:
\begin{equation}\label{generalized}
{{\tilde\gamma_{ij}:=\frac{\gamma_{ij}}{\sqrt[3]{\gamma /f}},\qquad
\tilde\pi^{ij}:=\sqrt[3]{\frac{\gamma}{f}}\left(\pi^{ij}-\frac{1}{3}\pi\gamma^{ij}\right)},}
\end{equation}
where additionally to the determinant $\gamma$ defined in (\ref{gammadet}),
the determinant of background metric $f$ is appeared:
$$f:=\det (f_{ij}).$$
The conformal metric $\tilde\gamma_{ij}$ (\ref{generalized}) is a tensor field, {\it id est}
it transforms according to the tensor representation of the group of diffeomorphisms.
The scaling variable $(\gamma/f)$ is a scalar field,
{\it id est} it is an invariant relative to diffeomorphisms.

We add to the conformal variables (\ref{generalized}) a canonical pair:
{\it a local intrinsic time} $D$ and a Hamiltonian density $\pi_D$
by the following way
\begin{equation}\label{DiracTpi}
{D:=-\frac{1}{3}\ln\left(\frac{\gamma}{f}\right),\qquad \pi_D:=\pi .}
\end{equation}
The formulae (\ref{generalized}), (\ref{DiracTpi}) define {\it the scaled Dirac's mapping} as a
mapping of the fiber bundles
\begin{equation}\label{generalizedD}
(\gamma_{ij}, \pi^{ij})\mapsto (D,\pi_D; \tilde\gamma_{ij}, \tilde\pi^{ij}).
\end{equation}
Riemannian superspaces of metrics $({}^3M)$ are defined on a compact Hausdorff manifolds $\Sigma_t$.
Denote the set, each point of which presents all various Riemannian metrics as ${\rm Riem}({}^3M)$.
Since the same Riemannian metric can be written in different coordinate systems,
we identify all the points associated with coordinate transformations of the diffeomorphism group
${\rm Diff}({}^3M)$.
All points receiving from some one by coordinate transformations of the group are called its orbit.
By this way the WDW superspace is defined as coset:
$$
{}^{(3)}\mathfrak{G}:={\rm Riem}({}^3M)/{\rm Diff} ({}^3M).
$$
As the saying goes \cite{Baierlein,Vision}, the superspace is the arena of Geometrodynamics.
Denote by ${}^{(3)}\mathfrak{G}{}^*$ the space of corresponding canonically conjugated densities of momenta.
According to the scaled Dirac's mapping (\ref{generalizedD}), we have the functional mapping of WDW phase
superspace of metrics $\gamma_{ij}$ and corresponding densities of their momenta $\pi^{ij}$
to WDW conformal superspace of metrics $\tilde\gamma_{ij}$, densities of momenta
$\tilde\pi^{ij}$; the local intrinsic time $D$, and the Hamiltonian density $\pi_D$:
$${{}^{(3)}\mathfrak{G}\times {}^{(3)}\mathfrak{G}{}^*
\to {}^{(3)}\mathfrak{\tilde{G}}\times {}^{(3)}\mathfrak{\tilde{G}}{}^*.}$$
We can conclude that the WDW phase superspace
$${}^{(3)}\mathfrak{\tilde{G}}\times {}^{(3)}\mathfrak{\tilde{G}}{}^*$$
is extended one, if we draw an analogy with relativistic mechanics \cite{Lanczos}.

York constructed the so-called extrinsic time \cite{York} as a trace of the extrinsic curvature tensor. Hence,
it is a scalar, and such a definition since that time is legalized in the theory of gravitation \cite{MTW}.
We built the intrinsic time with use of a ratio of the determinants of spatial metric tensors.
So to speak, the variables of the canonical pair: time - Hamiltonian density in the extended phase space,
in opposite to York's pair, are reversed.

Now the canonical variables are acquired a clear physical meaning.
The local time $D$ is constructed in accordance with the conditions imposed on the internal time.
The spatial metric $\gamma_{ij}$ carries information about the inner time.
According to the scaled Dirac's mapping (\ref{generalized}), (\ref{DiracTpi}),
the metric $\gamma_{ij}$ is factorized in the exponent function of the inner time $D$ and the conformal
metric $\tilde\gamma_{ij}$. So, the extracted intrinsic time has the spatial geometric origin.
Unlike to homogeneous examples, the intrinsic time $D$, in generic case, is local so-called many-fingered time.

The Poisson brackets (\ref{PB}) between new variables are the following
\begin{eqnarray}
\{D(t,{\bf x}), \pi_D(t,{\bf x}')\}&=&-\delta ({\bf x}-{\bf x}'),\nonumber\\
\{\tilde\gamma_{ij}(t,{\bf x}),\tilde\pi^{kl}(t,{\bf x}')\}&=&\tilde\delta_{ij}^{kl}\delta ({\bf x}-{\bf x}'),
\nonumber\\
\{\tilde\pi^{ij}(t,{\bf x}),\tilde\pi^{kl}(t,{\bf x}')\}&=&\frac{1}{3}(\tilde\gamma^{kl}\tilde\pi^{ij}-
\tilde\gamma^{ij}\tilde\pi^{kl}) \delta ({\bf x}-{\bf x}')\nonumber,
\end{eqnarray}
where
$$\tilde\delta_{ij}^{kl}:=\delta_i^k\delta_j^l+\delta_i^l\delta_j^k-
\frac{1}{3}\tilde\gamma^{kl}\tilde\gamma_{ij}$$
is the conformal Kronecker delta function  with properties:
$$\tilde\delta^{ij}_{ij}=5,~~~~~\quad \tilde\delta_{kl}^{ij}\tilde\delta_{mn}^{kl}=\tilde\delta_{mn}^{ij},$$
$$\tilde\delta_{ij}^{kl}\tilde\gamma_{kl}=\tilde\delta_{ij}^{kl}\tilde\gamma^{ij}=0,\quad~~~~~
\tilde\delta_{ij}^{kl}\tilde\pi^{ij}=\tilde\pi^{kl}.$$
The matrix $\tilde\gamma^{ij}$ is the inverse conformal metric, {\it id est}
$$\tilde\gamma^{ij}\tilde\gamma_{jk}=\delta^i_k,\qquad
\tilde\gamma^{ij}=\sqrt[3]{\frac{\gamma}{f}}\gamma^{ij}.$$
There are only five independent pairs $(\tilde\gamma_{ij}, \tilde\pi^{ij})$ per space point,
because of the properties
$$\tilde\gamma:=\det||\tilde\gamma_{ij}||=f,\qquad \tilde\pi:=\tilde\gamma_{ij}\tilde\pi^{ij}=0.$$
The generators $(D({\bf x},t)$, $\pi_D ({\bf x},t))$ form the subalgebra of the nonlinear Lie algebra.

Let us notice, that the problem of time and conserved dynamical quantities does not exist
in asymptotically flat worlds \cite{Regge}.
Time is measured by clocks of observers located at a sufficient far distance from the gravitational
objects. For this case the super-Hamiltonian (\ref{superHam}), constructed of the constraints,
is supplemented additionally by surface integrals at infinity.
Therefore, we focus our attention in the present paper on the cosmological problems only.

The Einstein's theory of gravitation is obtained from the shape dynamics by fixing the
St$\ddot{\rm u}$ckelberg's field (\ref{fixshape}):
\be
e^{-4\hat\phi}=\sqrt[3]{\frac{\gamma}{f}},
\ee
so that the factor is equal to
\be
\sqrt{\frac{f}{\gamma}}=\frac{e^{6\phi}}{<e^{6\phi}>_\gamma}.
\ee
The reparametrization constraints of shape dynamics are:
\bea
&&{\cal H}=\int_\Sigma\,d^3x\sqrt\gamma\left(e^{6\hat\phi [\gamma, \pi, x)}-1\right),\nonumber\\
&&{\cal H}^i=-2\nabla_j\pi^{ij},\quad {\cal Q}=4(\pi-<\pi>\sqrt\gamma).\nonumber
\eea
Here $e^{6\hat\phi [\gamma, \pi, x)}$ is the solution of the Lichnerowicz -- York equation.

\section{Deparametrization}

For making deparametrization we utilize the substitution (\ref{generalized}) and obtain
\bea
\tilde\pi^{ij}\dot{\tilde\gamma}_{ij}&=&\left(\pi^{ij}-\frac{1}{3}\pi\gamma^{ij}\right)\dot\gamma_{ij}+\nonumber\\
&+&\sqrt[3]\gamma\left(\pi^{ij}-\frac{1}{3}\pi\gamma^{ij}\right)\gamma_{ij}\left(\gamma^{-1/3}\right)^{.}=\nonumber\\
&=&\pi^{ij}\dot\gamma_{ij}-\frac{1}{3}\pi\left(\ln\gamma\right)^{.}.\nonumber
\eea

Then, after using the variables (\ref{DiracTpi}), we get
$$
\pi^{ij}\frac{d}{dt}\gamma_{ij}=\tilde\pi^{ij}\frac{d}{dt}\tilde\gamma_{ij}-\pi_D\frac{d}{dt}D.
$$
Then ADM functional of action (\ref{ADMvar}) takes the form
\bea
W&=&\int\limits_{t_I}^{t_0} dt\int\limits_{\Sigma_t} d^3x\left[
\left(\tilde\pi^{ij}_L+\tilde\pi^{ij}_{TT}\right)\frac{d}{dt}\tilde\gamma_{ij}-
\pi_D\frac{d}{dt}D-\right.\nonumber\\
&-&\left.N{\cal H}_\bot-N^i{\cal H}_i\right],\label{actionHilbert}
\eea
where the conformal momentum densities are decomposed on longitudinal and traceless - transverse parts,
the momentum is expressed through the scalar of extrinsic curvature:
$$
\tilde\pi^{ij}:=\tilde\pi^{ij}_L+\tilde\pi^{ij}_{TT},\qquad K=\frac{\pi}{2\sqrt\gamma}.
$$
The momentum $\pi_D$ is expressed out of the Hamiltonian constraint (\ref{H})
\bea
&&\pi_D [\tilde\pi^{ij}_L, \tilde\pi^{ij}_{TT},\tilde\gamma_{ij}, D]=\label{pfound}\\
&&=\sqrt{6\gamma}
\left[8\Psi^5\tilde\Delta\Psi
-\tilde{R}\Psi^4+\tilde\pi_{ij}\tilde\pi^{ij}+
\tilde{T}_{\bot\bot}\right]^{1/2},\nonumber
\eea
where
$$\Psi=\left(\frac{\gamma}{f}\right)^{1/12}.$$
The property
$$
\tilde{A}_{ij}\tilde{A}^{ij}=\tilde\pi_{ij}\tilde\pi^{ij}
$$
was utilized here.

Picking $D$ to be the time and setting an equality of the time intervals
$$dD=dt,$$
we can partially reduce the action (\ref{actionHilbert}).

Substituting the expressed $\pi_D$ (\ref{pfound})
into the ADM functional of action (\ref{actionHilbert}),
we get the functional presimplectic 1-form \cite{Olver,PTwo}  with local time $D$
\bea
\omega^1&=&\int\limits_{\Sigma_D}
d^3x\left[\left(\tilde\pi^{ij}_L+\tilde\pi^{ij}_{TT}\right)
d\tilde\gamma_{ij}-\right.\nonumber\\
&-&\left.\pi_D
\left(\tilde\pi^{ij}_L,\tilde\pi^{ij}_{TT},\tilde\gamma_{ij}, D\right)d D
\right].\label{omega1}
\eea

\section{Global time}

A global time exists in homogeneous cosmological models (see, for example, papers
\cite{Misner,Kasner,Kuchar,Isham,Pavlov,PavlovFlat,Pavlov1}).
For getting a global time in general  case, let us set the CMC gauge \cite{York}
(consideration especially for closed manifolds see \cite{JamesI}) on every slice, which labeled by the
coordinate time $t$,
\be\label{restrictions}
K\equiv -3\kappa=K(t),
\ee
where
$$\kappa:=\frac{1}{3}(\kappa_1+\kappa_2+\kappa_3)$$
is a mean curvature of the hypersurface $\Sigma_t$ -- arithmetic mean of the principal curvatures
$\kappa_1, \kappa_2, \kappa_3$.

Instead of Dirac's variables $(D, \pi_D)$ (\ref{hatTpT}) for obtaining global canonical conjugated variables
$$
\int_\Sigma\,d^3y D(y),\qquad
\int_\Sigma\,d^3x \pi_D (x)
$$
it is preferable to take York's transparent ones
\be\label{York}
T=-\sqrt\gamma,\qquad \pi_T=\frac{2\pi}{3\sqrt\gamma}.
\ee
Although, $T$ is a scalar density, we need not construct a scalar to get a global time.
Notice, that in generic case, a tensor density of weight $n$ is such a quantity $\tau$, that
$$\tau=\gamma^{n/2}T,$$
where $T$ is a tensor field.

A volume of a hypersurface
$$V:=\int_\Sigma d^3x\sqrt\gamma $$
plays a role of global time and a mean value of density of momentum
\be\label{reducedHam}
H :=\frac{2}{3}<\pi>=\frac{2}{3}\frac{\int_\Sigma d^3x \pi (x)}{\int_\Sigma d^3x\sqrt{\gamma (x)}}.
\ee
is a Hamiltonian.
Let us find the Poisson bracket between these nonlocal characteristics
$$
\left\{
\int_\Sigma\,d^3y\sqrt\gamma (y),\frac{2}{3}\frac{\int_\Sigma\,d^3y\pi (y)}{\int_\Sigma\,d^3y\sqrt\gamma (y)}
\right\}.
$$
We calculate functional derivatives of defined above functionals
$$
\frac{\delta}{\delta\gamma_{ij}(x)}{\int_\Sigma\,d^3y\sqrt\gamma (y)}=
\frac{1}{2}\sqrt\gamma (x)\gamma^{ij}(x).
$$
\bea
&&\frac{\delta}{\delta\pi^{ij}(x)}<\pi>=\frac{1}{V}\frac{\delta}{\delta\pi^{ij}(x)}\int_\Sigma\,d^3y\pi (y)=\nonumber\\
&=&\frac{1}{V}\frac{\delta}{\delta\pi^{ij}(x)}\int_\Sigma\,d^3y\pi^{ij}(y)\gamma_{ij}(y)=
\frac{1}{V}\gamma_{ij}(x).\nonumber
\eea
Hence,
\be
\{V, \frac{2}{3}<\pi>\}=\frac{1}{2V}\int_\Sigma\,d^3x\sqrt\gamma(x)\gamma^{ij}(x)\gamma_{ij}(x)=1.\nonumber
\ee
The corresponding Poisson bracket is canonical as in Ashtekar's approach \cite{Gorobey}.

Then the second term in (\ref{actionHilbert}) can be processed because of
$$
\frac{dD}{dt}=-\frac{1}{3\gamma}\frac{d\gamma}{dt}=-\frac{2}{3\sqrt\gamma}\frac{d\sqrt\gamma}{dt}.
$$
The York's condition (\ref{restrictions}) sets a slicing,
allowing to obtain a global time.

After Hamiltonian reduction and deparametrization procedures were executed, we yield the action
\bea\label{actionreduction}
&&W=\int\limits_{V_I}^{V_0} dV\int\limits_{\Sigma_t} d^3x\left[
\tilde\pi^{ij}\frac{d\tilde\gamma_{ij}}{d V}\right]-\\
&&-\int\limits_{V_I}^{V_0} H\, d V
-\int\limits_{V_I}^{V_0}\, dV\int\limits_{\Sigma_t} d^3x N^i {\cal H}_i
\nonumber
\eea
with the Hamiltonian (\ref{reducedHam}).

The Hamiltonian constraint is algebraic of the second order relative to $K$ that is
characteristic for relativistic theories. The coordinate time $t$ parameterizes the constrained theory.
The Hamiltonian constraint is a result of a gauge arbitrariness of the spacetime slicing
into space and time.

Let us notice, if the York's time was chosen \cite{KKuchar, Isenberg},
one should have to resolve the Hamiltonian constraint with respect to the variable $D$,
that looks unnatural difficult.
As was noted in \cite{Wald}, if the general relativity could be deparametrized,
a notion of total energy in a closed universe could well emerge. As follows from (\ref{pfound}),
in a generic case, the energy of the Universe is not conserved.

In CMC gauge we one yields
\be
<\pi>=\frac{2}{V}\int_\Sigma\, d^3x\sqrt\gamma K=2K.
\ee
The CMC gauge make condition to the lapse function $N$, if suppose the shift vector is zero $N_i=0$:
$$
K_{ij}=-\frac{1}{2N}\dot\gamma_{ij}.
$$
Then the trace is
$$K=\gamma^{ij}K_{ij}=-\frac{1}{2N}\gamma^{ij}\dot\gamma_{ij}=-\frac{1}{2N}\left(\ln\gamma\right)^{.}.$$

So, we find that the lapse function is not arbitrary, but defined according to the following restriction
\be
N=-\frac{1}{2K}\frac{d}{dt}\ln\gamma .
\ee

In a particular case, when $K={\rm const}$  for every hypersurface,
the Hamiltonian does not depend on time, so the energy of the system is conserved.
One maximal slice $(K=0)$ exists at the moment of time symmetry.
If additionally we were restricted our consideration conformal flat spaces, we should got set of waveless
solutions \cite{Isenberg}. Black holes, worm holes belong to this class of solution.

\section{Conclusions}

In the present paper we have demonstrated that in Geometrodynamics, the many-fingered intrinsic time is a
scalar field. For its construction a background metric was introduced.
To obtain reasonable dynamical characteristics,
in the capacity of background metric one should choose a suitable for corresponding topology a space metric.
For a generic case it is a tangent space, for asymptotically flat problems -- flat one \cite{Gor},
and for cosmological problems -- a compact, or non-compact corresponding manifold.
Hamiltonian approach to obtaining physical observables is not covariant unlike to Lagrangian one.
It seems quite natural: The Lagrangian approach is used to obtain an invariant functional of action,
covariant field equations, suitable for any frame of reference, and the Hamiltonian one --- for getting
physical observables for a given observer in his frame of reference.

The idea of introducing background fields is used traditionally under considering various theoretical problems.
Let us list here some well-known ones. In problems of studying vacuum polarization and quantum particle creation
on curved spaces background fields are necessary for extraction physical quantities \cite{Birrell}.
The procedures of regularization and renormalization of the Casimir vacuum energy are considered in
\cite{Bordag}. Renormalization involves comparing of some characteristics to obtain as a result of subtraction
the physical observables. For construction of cosmological perturbations theory as a background metric the
Friedmann -- Robertson --Walker one stands \cite{Mukhanov}, \cite{Durrer}.
The presence of the Minkowski spacetime, as is shown in \cite{Sol}, is necessary to obtain conserved quantities
of gravitational field. The background Minkowski spacetime is presented hidden in asymptotically flat space
problems \cite{ADM}. Topological Casimir energy of quantum fields in closed hyperbolic universes is calculated
in \cite{Muller}, \cite{Fagundes}.
The problem of obtaining of energy-momentum tensor of the gravitational field in Ricci-flat backgrounds is
discussed in \cite{Grishchuk}.

In Geometrodynamics, the many-fingered intrinsic time is a scalar field. After the York's gauge was implemented,
the deparametrization leads to the global time -- the value of the hypersurface of the Universe .
In application to the problem of the Universe,
the global time is a function of the FRW model scale. It is in agreement with
the stationary Einstein's conception of the Universe \cite{EinsteinCosmology}.
Thus we avoid an unpleasant unresolved problem yet of
initial singularity (Big Bang) in the Standard cosmology. The Friedmann equation has a sense of the formula,
connected time intervals (intrinsic, coordinate, conformal) \cite{PP, ZakhP, Pavlovexact}.
If we wish accept the York's extrinsic time,
we get the Friedmann equation as algebraic one. Hence, the connection between temporal intervals
(geometrical coordinate time
$t$ in the pseudo-Riemannian space and intrinsic one $D$ in the WDW superspace) is lost.
Instead of an expansion of the Universe (Standard cosmology) we accept the rate of mass (Conformal cosmology)
\cite{MY}.

In frame of Conformal cosmology it is meaningful to speak of the energy of the Universe that was lost
in Standard cosmology \cite{Wald}.
The some authors (see, for example, \cite{Beluardi, GomesKos, Valentini}) prefer utilizing the York time as a
real time and a volume of the Universe as an operator of evolution.
A linking theory that proves the equivalence of General Relativity and Shape Dynamics was constructed in
\cite{Link}.

{\it Nothing is more mysterious and elusive than time. it seems to be the most powerful force in the universe,
carrying us inexorably from birth to death. But what exactly is it? St. Augustine, who died in AD 430, summed
up the problem thus: `If nobody asks me, I know what time is, but if I am asked then I am at a loss what to say'.
All agree that time is associated with change, growth and decay, but is it more than this? Questions abound.
Does time move forward, bringing into being an ever-changing present? Does the past still exist?
Where is the past? Is the future already predetermined, sitting here waiting for us though we know not what
it is?} \cite{Barbour}.

Before $XX$-th century these questions belonged to philosophers. The Einstein's theory
of gravitations allows to stand these questions in physics frame.
The changing volume of the Universe in Standard Cosmology, or changing or masses of elementary particles in Conformal Cosmology is the measure of time, not time is the measure of change.
The so-called expansion of the Universe is able to be tied with time of the Universe.

\section*{Acknowledgment}

For fruitful discussions, comments, and criticisms, I would like to thank Prof. V.N. Pervushin.
I am grateful for discussions to participants of the seminars of the Institute for Gravitation and Cosmology,
the People's Friendship University of Russia (PFUR), Moscow,
and the Russian Gravitational Association (VNIIMS), Moscow.


\end{document}